\documentclass[epj]{webofc}
\usepackage[varg]{txfonts}   
\wocname{EPJ Web of Conferences}
\woctitle{CONF12}
\newcommand{\ts}[1]{\textrm{\tiny #1}}
\newcommand{\ms}[1]{\textrm{\tiny $#1$}}

\newcommand{\be}{\begin{equation}}
\newcommand{\ee}{\end{equation}}
\newcommand{\bse}{\begin{subequations}}
\newcommand{\ese}{\end{subequations}}
\newcommand{\ba}{\begin{eqnarray}}
\newcommand{\ea}{\end{eqnarray}}
\renewcommand{\(}{\left(}
\renewcommand{\)}{\right)}

\newcommand{\bea}{\begin{eqnarray}}
\newcommand{\eea}{\end{eqnarray}}

\newcommand{\mn}{\mu\nu}

\newcommand{\wE}{\hat{\mathcal{E}}}
\newcommand{\wP}{\hat{\mathcal{P}}}

\begin{document}
\selectlanguage{english}
\title{Semiholography for heavy ion collisions}
%
%

\author{Ayan Mukhopadhyay\inst{1,2}\fnsep\thanks{\email{ayan@hep.itp.tuwien.ac.at}} \and
        Florian Preis\inst{1}\fnsep\thanks{\email{fpreis@hep.itp.tuwien.ac.at}}}

\institute{Institut f\"ur theoretische Physik, Technische Universit\"at Wien, 1040 Vienna, Austria \and CERN, Theoretical Physics Department, 1211 Geneva 23, Switzerland}

\abstract{%
  The formation of QGP in heavy ion collisions gives us a great opportunity for learning about nonperturbative dynamics of QCD. Semiholography provides a new consistent framework to combine perturbative and non-perturbative effects in a coherent way and can be applied to obtain an effective description for heavy ion collisions. In particular, it allows us to include nonperturbative effects in existing glasma effective theory and QCD kinetic theory for the weakly coupled saturated degrees of freedom liberated by the collisions in the initial stages in a consistent manner. We argue why the full framework should be able to confront experiments with only a few phenomenological parameters and present feasibility tests for the necessary numerical computations. Furthermore, we discuss that semiholography leads to a new description of collective flow in the form of a generalised non-Newtonian fluid. We discuss some open questions which we hope to answer in the near future. }
\maketitle
\tableofcontents
\section{Introduction}
 
 There are very good reasons to believe that we need to combine both weakly coupled perturbative and strongly coupled nonperturbative effects to find an effective theory for the formation and evolution of the quark-gluon plasma in heavy ion collisions. 
 
 The initial stages of the collision can be understood using glasma effective theory which takes advantage of saturation physics \cite{Iancu:2000hn,Gelis:2010nm}. The gluons liberated by the collisions are mostly small-$x$ (slowly moving along the collision axis) partons of the nuclei which form a weakly coupled over-occupied (i.e. with occupation numbers of $\mathcal{O}(1/\alpha_s)$) system by virtue of which these can be described by classical Yang-Mills equations. The latter are sourced by colour charges of the large-$x$ (rapidly moving along the collision axis) gluons. For these gluons, $x> x_0$ with $x_0$ being a cut-off value of $x$. The evolution of their colour charge distribution (frozen on the time scale of collisions) with the cut-off $x_0$ can be followed via perturbative QCD \cite{Iancu:2000hn}. Typically the distribution is Gaussian with a transverse width $1/Q_s(x_0)$, where $Q_s$ is the so-called saturation scale which is much higher than the confinement scale \cite{Kovner:1995ts,Kovner:1995ja,Lappi:2006fp}. 
 
 At a slightly later stage, one cannot use classical Yang-Mills equations even for the small-$x$ gluons because of dilution due to expansion. Therefore one needs to interpolate classical Yang-Mills equations with kinetic theory. Remarkably, the key aspects of the transition from classical Yang-Mills equations to a kinetic description has been understood recently \cite{Kurkela:2011ti,Kurkela:2014tea,Kurkela:2015qoa}. 
 
Nevertheless, one cannot ignore strong coupling effects because soft gluons (with large transverse sizes) are also radiated and are expected to form a strongly coupled thermal bath. These could be responsible for remarkably fast transition to hydrodynamics with a very small shear-viscosity to entropy density ratio and other collective effects \cite{Policastro:2001yc,Back:2004je}. Therefore, to understand the phenomenology of the formation and evolution of the quark gluon plasma, we need to learn how to bring together various degrees of freedom at diverse energy scales in a coherent manner. This is preferable to doing either weakly coupled or strongly coupled calculations exclusively, and then interpolating in the coupling.
 
Semiholography is a framework for combining perturbative and nonperturbative effects in a consistent way to give a complete effective description of the dynamics at a wide range of energy scales. In the case of the quark-gluon plasma, fortunately many issues involved in the full construction can be handled with relative conceptual ease as will be described below. In this approach, the soft infrared gluons giving nonperturbative effects are assumed to have a dual holographic description in the form of an appropriate classical theory of gravity. As discussed below, one can argue that the full construction should involve only a few effective parameters. 

\section{Semiholography with the glasma}
Assuming that degrees of freedom at the confinement scale do not play any major role until hadronisation, we can describe the soft gluonic system as a strongly coupled holographic CFT (conformal field theory). The latter can be better described as an emergent strongly coupled holographic large $N$ Yang-Mills theory with an approximate conformal symmetry in the relevant range of energy scales and which models the nonperturbative sector. A simplistic holographic description in the form of Einstein's gravity in anti-de Sitter space minimally coupled to a massless dilaton and a massless axion (capturing all relevant deformations as discussed below) can be expected to work reasonably well.

The semiholographic model combining the glasma and the soft sector (with appropriate initial conditions given by glasma effective theory) can then be written in the form \cite{Iancu:2014ava,Mukhopadhyay:2015smb}:
\begin{equation}
S = S_{\rm YM}[A_\mu^a] + W^{\rm CFT}\Big[\tilde{g}_{\mu\nu}[A_\mu^a], \delta\tilde{g}_{\rm YM}[A_\mu^a], \tilde{\theta}[A_\mu^a]\Big].
\end{equation}
Above, $W^{\rm CFT}$ is the generating functional of connected correlation functions of the strongly coupled holographic CFT modelling the infrared sector with sources that are functionals of the perturbative glasma fields, and
\begin{equation}
S_{\rm YM}[A_\mu^a] = - \frac{1}{4N_c}\int {\rm d}^4 x \, \, {\rm tr} \left(F_{\alpha\beta} F^{\alpha\beta}\right).
\end{equation}
Furthermore, the holographic CFT (strictly speaking we assume approximate conformal invariance only) by virtue of our assumption of being an emergent Yang-Mills theory can undergo only \textit{three} marginal deformations involving the change in the effective metric, its Yang-Mills coupling and its theta parameter. The background metric becomes effectively $\tilde{g}_{\mu\nu}$, the Yang-Mills coupling changes from infinity by $\delta\tilde{g}_{\rm YM}$ and the theta parameter changes from zero to $\tilde{\theta}_{\rm YM}$ due to the influence of perturbative fields. By virtue of holographic duality,
\begin{equation}
W^{\rm CFT}\Big[\tilde{g}_{\mu\nu}[A_\mu^a], \delta\tilde{g}_{\rm YM}[A_\mu^a], \tilde{\theta}_{\rm YM}[A_\mu^a]\Big] = S^{\rm on-shell}_{grav}[\tilde{g}_{\mu\nu}= g^{\rm (b)}_{\mu\nu}, \delta\tilde{g}_{\rm YM} = \phi^{(b)}, \tilde{\theta} = \chi^{\rm (b)}],
\end{equation}
i.e. $W^{\rm CFT}$ can be identified with the on-shell gravitational action of the dual classical gravity theory. Furthermore, $\tilde{g}_{\mu\nu}$ is identified with the \textit{boundary metric} $g^{\rm (b)}_{\mu\nu}$ giving the leading asymptotic behaviour of the bulk metric, $\delta\tilde{g}_{\rm YM}$ is identified with the boundary value $\phi^{\rm (b)}$ of the bulk dilaton $\Phi$ and $\tilde{\theta}$ is identified with the boundary value $\chi^{\rm (b)}$ of the bulk axion $\mathcal{X}$.  We can thus write the full semiholographic action in the form:
\begin{equation}\label{modified-glasma}
S = S_{\rm YM}[A_\mu^a] + S^{\rm on-shell}_{grav}\Big[g^{\rm (b)}_{\mu\nu}[A_\mu^a], \phi^{\rm (b)}[A_\mu^a], \chi^{\rm (b)}[A_\mu^a]\Big].
\end{equation}

Finally, we specify that:
\begin{eqnarray}
g^{\rm (b)}_{\mu\nu} &=&\eta_{\mu\nu} + \frac{\gamma}{Q_s^4} t^{\rm cl}_{\mu\nu}, \, \, \, \,{\rm with}\,\,\,\, t^{\rm cl}_{\mu\nu} = \frac{1}{N_c} {\rm tr} \left(F_{\mu\alpha}F_\nu^{\phantom{\mu}\alpha} - \frac{1}{4}\eta_{\mu\nu}F_{\alpha\beta}F^{\alpha\beta}\right),\label{source-1} \\
\phi^{\rm (b)} &=& \frac{\beta}{Q_s^4}h^{\rm cl}, \, \, \, \,{\rm with}\,\,\,\, h^{\rm cl} = \frac{1}{4N_c} {\rm tr}\left(F_{\alpha\beta}F^{\alpha\beta}\right),\label{source-2}\\
\chi^{\rm (b)} &=& \frac{\alpha}{Q_s^4}a^{\rm cl}, \, \, \, \,{\rm with}\,\,\,\, a^{\rm cl} = \frac{1}{4N_c} {\rm tr}\left(F_{\alpha\beta}\tilde{F}^{\alpha\beta}\right)\label{source-3}.
\end{eqnarray}
Above $ t^{\rm cl}_{\mu\nu}$ is thus the perturbative energy-momentum tensor, $h^{\rm cl}$ is the perturbative Lagrangian density and $a^{\rm cl}$ is the perturbative Pontryagin density. The suffix $\rm 'cl'$ indicates that these are functionals of the classical YM fields of the glasma. Since $Q_s$ is the energy scale of the hard part set by the initial conditions, it should provide the scales of hard-soft interactions naturally.

It can be readily shown that in consistency with the variational principle the modified classical glasma action (\ref{modified-glasma}) can also be written in the form:
\begin{equation}\label{modified-glasma-2}
S = S_{\rm YM}[A_\mu^a] +\frac{1}{2} \int {\rm d}^4x \, \overline{\mathcal{T}}^{\mu\nu}g^{(b)}_{\mu\nu} + \int {\rm d}^4x \, \overline{\mathcal{H}}\phi^{\rm (b)} + \int {\rm d}^4x \, \overline{\mathcal{A}}\chi^{(b)},
\end{equation}
where 
\begin{eqnarray}
\overline{\mathcal{T}}^{\mu\nu} = 2 \frac{\delta S^{\rm on-shell}_{grav}}{\delta g^{\rm (b)}_{\mu\nu}}, \quad \overline{\mathcal{H}} = \frac{\delta S^{\rm on-shell}_{grav}}{\delta \phi^{\rm (b)}}, \quad \overline{\mathcal{A}} = \frac{\delta S^{\rm on-shell}_{grav}}{\delta \chi^{\rm (b)}},
\end{eqnarray}
with the right hand sides of the above equations evaluated at the values given by (\ref{source-1}, \ref{source-2}, \ref{source-3}).

\textit{The full dynamics needs to be solved self-consistently in an iterative fashion.} The action in the form (\ref{modified-glasma-2}) clearly shows that the glasma equations are modified by holographic operators that appear as self-consistent mean fields. Furthermore, it also evident from (\ref{modified-glasma-2}) that the perturbative sector is deformed by the nonperturbative sector only in a marginal way (via dimension $4$ operators only), although one of the deformations is tensorial and the couplings are functionals of the self-consistent expectation values of the operators of the non-perturbative sector. As noted before, the nonperturbative sector described by the dual classical gravity theory is also deformed marginally by the perturbative sector in a reciprocative manner. Therefore, \textit{eventually when we include quantum effects in the glasma and those in the dual gravity description, both of them will be solvable at each step of iteration (without the need for introducing extra parameters for renormalisation) by virtue of marginal deformations of each sector. This explains why we should have an approximately good description of the QGP at large $N_c$ using only three additional hard-soft couplings $\alpha$, $\beta$ and $\gamma$.} Of course, all these couplings at a given scale $\Lambda$ should be a function of $\Lambda _{\rm QCD}/\Lambda$ such that they vanish in the UV (in the $\Lambda \rightarrow \infty$ limit) in which the physics should be mostly perturbative. The latter functions should be derived from the Lagrangian of QCD from first principles. We leave this task for the future. At present we just work with $\alpha$, $\beta$ and $\gamma$ at the scale $Q_s$ that is set by the initial conditions while treating them as phenomenological parameters.

\section{Solving the full dynamics iteratively}
Before we describe the iterative procedure of solving the dynamics, let us first derive the equation of motion for the glasma fields from the action (\ref{modified-glasma}) or equivalently (\ref{modified-glasma-2}). In either case we can use the chain rule, as for instance:
\begin{equation}
\frac{\delta S^{\rm on-shell}_{grav}}{\delta A_\mu^a(x)} = \int {\rm d}^4y \, \left(\frac{\delta S^{\rm on-shell}_{grav}}{\delta g^{\rm(b)}_{\mu\nu}(y)}\frac{\delta g^{\rm(b)}_{\mu\nu}(y)}{\delta A_\mu^a} + \cdots \right).
\end{equation}
If we derive from the form (\ref{modified-glasma-2}), we must treat the holographic operators $\overline{\mathcal{T}}^{\mu\nu}$, etc to be independent of the glasma fields $A_\mu^a$. In either way, we can arrive at the modified glasma equations (suppressing color indices):\footnote{We have assumed that the full system lives in flat Minkowski space $\eta_{\mu\nu}$. If this is not the case, slightly different covariant tensorial objects appear which are e.g. $\hat{\mathcal{H}} = \frac{1}{\sqrt{-g}}\overline{\mathcal {H}}$ with $g_{\mu\nu}$ being the fixed background metric for all the degrees of freedom. It can be readily seen that $\hat{\mathcal{H}}$ etc. transform in a covariant way under diffeomorphisms. When the background metric is $\eta_{\mu\nu}$, $\hat{\mathcal{H}}$ coincides with $\overline{\mathcal {H}}$. }
\begin{equation}\label{glasma-eom}
D_\mu F^{\mu\nu} = \frac{\beta}{Q_s^4}D_\mu \left(\overline{\mathcal{H}}F^{\mu\nu}\right)+ \frac{\alpha}{Q_s^4}\left(\partial_\mu \overline{\mathcal{A}}\right) F^{\mu\nu} + \frac{\gamma}{Q_s^4} D_\mu\left(\overline{\mathcal{T}}^{\mu\alpha}F_\alpha^{\phantom{\alpha}\nu}-\overline{\mathcal{T}}^{\nu\alpha}F_\alpha^{\phantom{\alpha}\mu}- \frac{1}{2}\overline{\mathcal{T}}^{\alpha\beta}\eta_{\alpha\beta}F^{\mu\nu}\right),
\end{equation}
with $D_\mu$ being the gauge-covariant derivative.

At each step of the iteration, $\overline{\mathcal{T}}^{\mu\alpha}$, etc. can be extracted from the classical gravity equations as follows. Firstly, we note that:
\begin{equation}\label{relation}
\overline{\mathcal{T}}^{\mu\nu} = \sqrt{-g^{\rm (b)}} \mathcal{T}^{\mu\nu}, \,\,\,\, {\rm with} \,\,\,\, \mathcal{T}^{\mu\nu} = \frac{2}{\sqrt{-g^{\rm (b)}}} \frac{\delta S^{\rm on-shell}_{grav}}{\delta g^{\rm (b)}_{\mu\nu}}, \,\, \,{\rm etc.}
\end{equation}
Then $\mathcal{T}^{\mu\nu}$ can be evaluated using the standard holographic dictionary. We need to solve the $5-$dimensional classical gravity equations with appropriate initial conditions and boundary metric and sources specified by (\ref{source-1}, \ref{source-2}, \ref{source-3}). We obtain a unique solution. Let $r$ be the radial direction such that $r=0$ is the boundary. Then we can extract $\mathcal{T}^{\mu\nu}$ from the asymptotic expansion of the $5-$dimensional metric $G_{MN}$ in the Fefferman-Graham coordinates which reads
\begin{eqnarray}
G_{rr} = \frac{l^2}{r^2}, \quad G_{r\mu} = 0, \\\nonumber
G_{\mu\nu} = \frac{l^2}{r^2} \left(g^{\rm(b)}_{\mu\nu} + \cdots +r^4\left( \frac{4\pi G_5}{l^3}\mathcal{T}_{\mu\nu}+ X_{\mu\nu}\right) + \cdots\right),
\end{eqnarray}
where $X_{\mu\nu}$ is an explicitly known local functional of $g^{\rm(b)}_{\mu\nu}$ \cite{deHaro:2000vlm}. Finally, we should use
\begin{equation}
\mathcal{T}^{\mu\nu} = g^{\rm (b) \mu\alpha}\mathcal{T}_{\alpha\beta}g^{\rm(b) \beta \nu},
\end{equation}
and then (\ref{relation}) to obtain $\overline{\mathcal{T}}^{\mu\nu}$. Similarly, we obtain $\overline{\mathcal{H}}$ and $\overline{\mathcal{A}}$ from the asymptotic expansions of $\Phi$ and $\mathcal{X}$ in the gravitational solution with specified boundary sources (\ref{source-1}, \ref{source-2}, \ref{source-3}). 

The iterative process of solution is as follows \cite{Iancu:2014ava}.
\begin{enumerate}
\item We first solve the glasma equations (\ref{glasma-eom}) with $\alpha = \beta = \gamma = 0$.
\item From this solution, we extract the sources (\ref{source-1}, \ref{source-2}, \ref{source-3}) for the gravitational problem solving which we obtain $\overline{\mathcal{T}}^{\mu\nu}$, $\overline{\mathcal{H}}$ and $\overline{\mathcal{A}}$.
\item We solve the glasma equations (\ref{glasma-eom}) again after inserting the above $\overline{\mathcal{T}}^{\mu\nu}$, $\overline{\mathcal{H}}$ and $\overline{\mathcal{A}}$ in them.
\item We go back to step 2 to solve the gravitational equations again with sources now specified by the new solution to the glasma equations. 
\item We then continue to repeat steps 3 and 2 successively until both the gravitational solution and the glasma fields converge to their final forms.
\end{enumerate}
At each step of the iteration, we hold initial conditions fixed. For the glasma, the initial conditions can be shown to be unaltered, i.e. given by the usual perturbative large-$x$ partonic color sources. For the gravitational part, the initial conditions should be that of pure AdS with vanishing dilaton and axion fields, indicating that the soft sector does not play any role in the initial stages.\footnote{In reality, we need to introduce a small technical complication. We need to put the initial condition for the gravitational part slightly before that of the perturbative glasma sector.} The gravitational sector can be solved using the method of characteristics \cite{Chesler:2008hg}.

\textit{A crucial consistency check of the full framework is that one can prove that there exists a local 
energy-momentum tensor $T^{\mu\nu}$ for the combined system that can be explicitly constructed and which is conserved in the background (flat Minkowski) metric (i.e. satisfying $\partial_\mu T^{\mu\nu} = 0$) in which all the degrees of freedom live \cite{Mukhopadhyay:2015smb}.} This energy-momentum tensor can be obtained from the action (\ref{modified-glasma}) or (\ref{modified-glasma-2}) by using the variational principle again. Its conservation serves as a check of convergence of the iterative process of obtaining the full dynamical solution. Explicitly,\footnote{Once again, if the background metric is not $\eta_{\mu\nu}$, we must replace $\overline{\mathcal{T}}^{\mu\nu}$ by $\hat{\mathcal{T}}^{\mu\nu}$, etc. When the background metric is $\eta_{\mu\nu}$, they coincide.}
\begin{eqnarray}\label{full-em}
T^{\mu\nu} &= &t^{\mu\nu} + \overline{\mathcal {T}}^{\mu\nu} \nonumber\\
&& -\frac{\gamma}{Q_s^4N_c} \overline{\mathcal {T}}^{\alpha\beta}\left({\rm tr}\left(F_\alpha^{\phantom{\alpha}\mu}F_\beta^{\phantom{\beta}\nu}\right)- \frac{1}{4}\eta_{\alpha\beta}{\rm tr}\left(F^{\mu\rho}F^\nu_{\phantom{\nu}\rho}\right)+\frac{1}{4}\delta^\mu_{(\alpha}\delta^\nu_{\beta)} {\rm tr}\left(F_{\alpha\beta}F^{\alpha\beta}\right)\right)\nonumber\\
&&-\frac{\beta}{Q_s^4 N_c}\overline{\mathcal{H}}\,{\rm tr}\left(F^{\mu\rho}F^\nu_{\phantom{\nu}\rho}\right)-\frac{\alpha}{Q_s^4}\overline{\mathcal{A}}\, a.
\end{eqnarray}
\section{Toy example}
\label{sec-1}

In this section we will restrict our model to a simple scenario to perform a numerical test for convergence of the iterative method of solving the full semiholographic dynamics \cite{Mukhopadhyay:2015smb}. We will only allow for a finite value of the coupling $\gamma$ and demand that the UV and the IR degrees of freedom are both spatially homogeneous and isotropic. For simplicity we choose the gauge group of the classical Yang--Mills theory to be $SU(2)$. In temporal gauge $A_t^a=0$ with the ansatz $A_i^a=f(t)\delta_i^a$, where $i$ denotes the spatial directions and $a$ denotes the SU(2) indices, we find $t_{\mu\nu}=p(t)\,\, {\rm diag}(3,1,1,1)$ with
\be
p(t)=\frac{1}{2}\left[f'(t)^2+f(t)^4\right],
\ee
being the YM pressure.

As we have set $\alpha = \beta = 0$ for simplicity, the only deformation of the IR-CFT involves the effective metric as designed by the glasma fields following (\ref{source-1}). The imposed symmetries of homogeneity and isotropy leads to a conformally flat $g^{\ts{(b)}}_{\mn}$. The dual holographic geometry should be a solution of pure Einstein's gravity as setting $\alpha = \beta = 0$ implies that the bulk dilaton and axion fields vanish too. Since the boundary metric is conformally flat, homogeneity and isotropy imply that the bulk solution should be a time-dependent diffeomorphism of the anti-de Sitter Schwarzschild black brane by virtue of Birkhoff's theorem. The latter is simply the dual of a thermal state of the IR-CFT. Furthermore, as bulk diffeomorphisms lift to a combination of diffeomeorphisms and conformal transformations at the boundary, $\overline{\mathcal{T}}^{\mu\nu}$ can be obtained from a time dependent coordinate plus conformal transformation that takes the thermal energy-momentum tensor of the IR-CFT in flat space (which takes the form $\mathcal{T}^{\mu\nu}=a_1 c\ {\rm diag}(3,1,1,1)$, where the constant $c$ sets the temperature dual to the black hole mass) to the energy-momentum tensor in the background metric designed by the homogeneous and isotropic glasma fields. It turns out that the necessary conformal transformation is given by the Weyl factor 
\be
\Omega(t)=\sqrt{1+\gamma p(t)/Q_s^4},
\ee 
while the necessary coordinate transformation of covariant objects is described by the matrix 
\be
\Lambda^\mu_{\ \nu}={\rm diag}\left(\sqrt{1-3\gamma p(t)/Q_s^4)}/\sqrt{1+\gamma p(t)/Q_s^4},1,1,1\right). 
\ee
The full result then consists of a contribution $\mathcal{T}^{{\rm (cov)}\mu\nu}$ transforming covariantly and an anomalous contribution $\mathcal{T}^{{\rm (an)}\mu\nu}$ given by \cite{PhysRevD.16.1712,Herzog:2013ed}. Thus
\begin{eqnarray}
\mathcal{T}^{\mu\nu} &=&\mathcal{T}^{{\rm (cov)}\mu\nu} +\mathcal{T}^{{\rm (an)}\mu\nu}, \quad {\rm with} \\ \mathcal{T}^{{\rm (cov)}\mu\nu} &=& a_1c\Omega^{-6}\ {\rm diag}\left[3(\Lambda^t_{\ t})^{-1},1,1,1\right] \quad {\rm and}\\
\mathcal{T}^{{\rm (an)}\mu\nu} &=& -\frac{a_4}{(4\pi)^2}\left[g^{\ts{(b)}\ \mu\nu}\(\frac{R^2}{2}-R_{\alpha\beta}R^{\alpha\beta} \)+ 2R^{\mu\lambda}R^\nu_{\phantom{\mu}\lambda} - \frac{4}{3}RR^{\mu\nu}\right].\label{eq:Tanom}
\end{eqnarray}
The Riemann curvature tensor above refers to that of the metric $g^{\ts{(b)}}_{\mu\nu}$. Note that for strongly coupled large $N_c$ holographic CFTs one finds $a_1=N_c^2/8\pi^2$ and $a_4=N_c^2/4$.

\textit{The high degree of symmetry in our set-up implies that the gravitational solution is a diffeomorphism of a pre-existing black hole solution. Since no entropy production is involved, the energy transfer between the hard and soft sectors should be reversible and this should be reflected in the oscillatory nature of the respective energy densities. Although our toy example provides a good testing ground of numerical convergence of the iterative procedure of solving the full dynamics proposed in \cite{Iancu:2014ava} and discussed above, it cannot be used to demonstrate thermalisation which is expected to arise for more general initial conditions.} The logical next step towards a more interesting case regarding the issue of thermalisation is either to incorporate anisotropy or to allow for matter degrees of freedom, by having $\beta\neq0$. Both of these additional complications include this particularly simple setting as a limiting case, thus it is also worthwhile to study it in great detail for future reference.

Even in our simple scenario the equations of motion for the gauge field $A_i^a=f(t)\delta_i^a$ is highly non-trivial and is given by:
\begin{equation}
f''(t)+2  \frac{1-\frac{1}{2} \frac{\gamma}{Q_s^4}(\wE +\wP )}{1+\frac{1}{2} \frac{\gamma}{Q_s^4}(\wE +\wP )}f(t)^3+\frac{1}{2}  \frac{\gamma}{Q_s^4}\frac{(\wE +\wP )'}{1+\frac{1}{2} \frac{\gamma}{Q_s^4}(\wE +\wP )} f'(t)=0,\label{eq:EOMiso}
\end{equation}
where $\wE:=\sqrt{-g^{\ts{(b)}}}\mathcal{T}^{tt}$ and $\wP:=\sqrt{-g^{\ts{(b)}}}\mathcal{T}^{xx}=\sqrt{-g^{\ts{(b)}}}\mathcal{T}^{yy}=\sqrt{-g^{\ts{(b)}}}\mathcal{T}^{zz}$. Note that the Ricci tensor and curvature scalar contributing to the anomalous part IR energy momentum tensor \eqref{eq:Tanom}, involve second order derivatives of the scaling factor $\Omega(t)$, and thus involve third order derivatives of the gauge field. Because of the last term on the left hand side of \eqref{eq:EOMiso}, the equation of motion for the gauge field is a fourth order ordinary differential equation (ODE), whereas in the limit of vanishing coupling $\gamma$ it is of second order. As discussed before, in the first step of iteration we set $\gamma = 0$ and solve (\ref{eq:EOMiso}) with initial conditions $f'(0) = 0$ which results in the fulfilment of  the Gauss Law constraint and $f(0) =(2p_0)^{1/4}$ chosen to set a desired value $p_0$ of the initial pressure in the hard (YM) sector. We then use this solution for $f(t)$ to determine $\wE$ and $\wP$ which have been so far set to a static thermal value. We continue the iteration until we find convergence. \textit{A rigorous test of convergence is provided by the conservation of energy-momentum tensor (\ref{full-em}) of the combined system in flat space (which in our simple scenario amounts to conservation of the total energy) to a desirable degree of numerical accuracy.}

\begin{figure}[h]
\centering
\sidecaption
\includegraphics[width=10cm,clip]{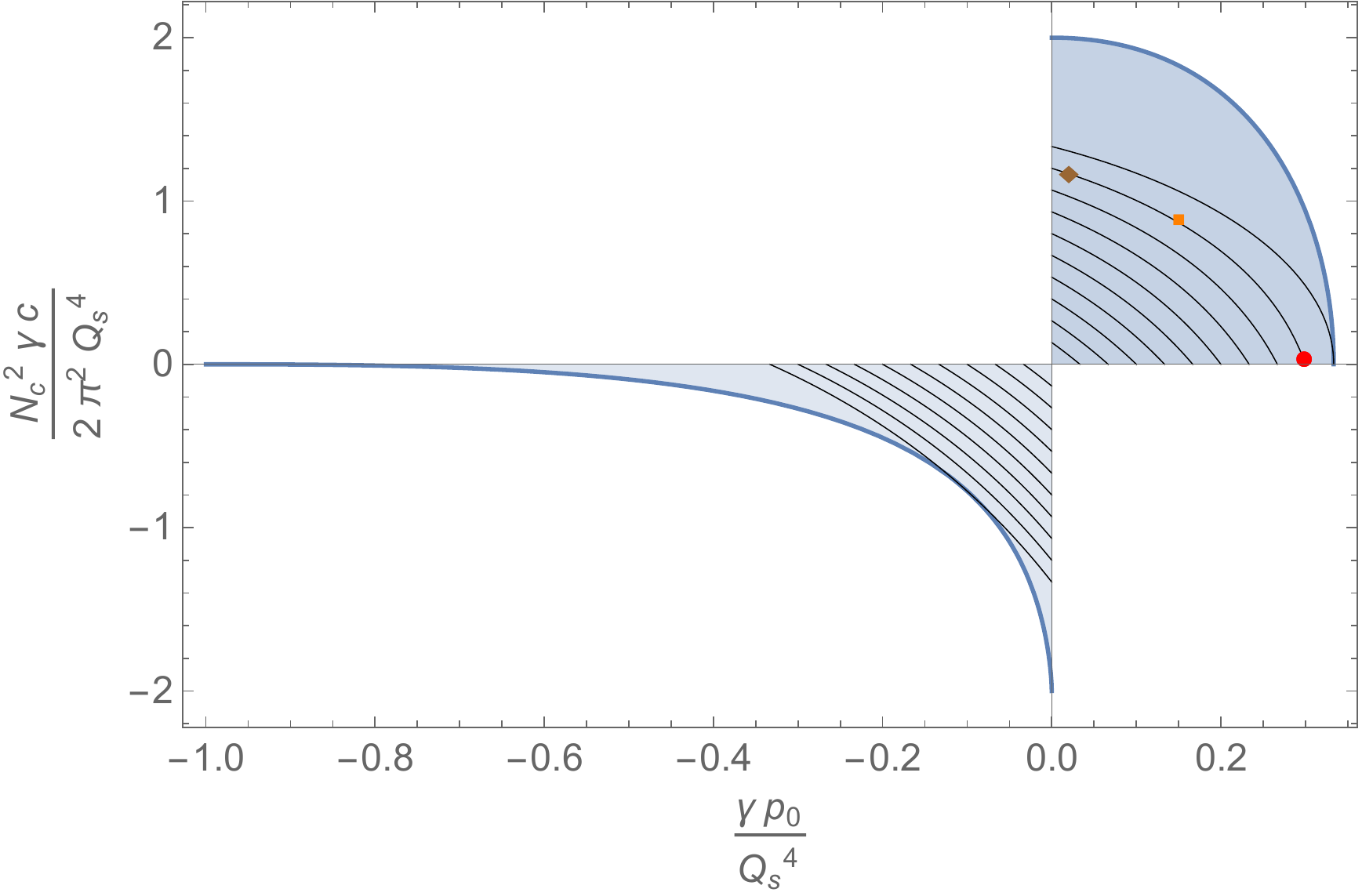}
\caption{The allowed (shaded) regions for the combination of parameters $\gamma p_0/Q_s^4$ and $N_c^2\gamma c/(2\pi^2Q_s^4)$ in the homogeneous and isotropic toy example with $\alpha=\beta=0$ such that regular solutions exist. The thin lines represent curves of constant $\gamma E/Q_s^4$. The point, square and diamond all lie on the $\gamma E/Q_s^4=0.9$ curve and mark the choices (a) $p_0=1.49\ Q_s^4$, (b) $p_0=0.75\ Q_s^4$ and (c) $p_0=0.1\ Q_s^4$ with $\gamma=0.2$ respectively.}
\label{fig:gammabounds}       
\end{figure}

\begin{figure}[h]
\centering
\includegraphics[width=7cm,clip]{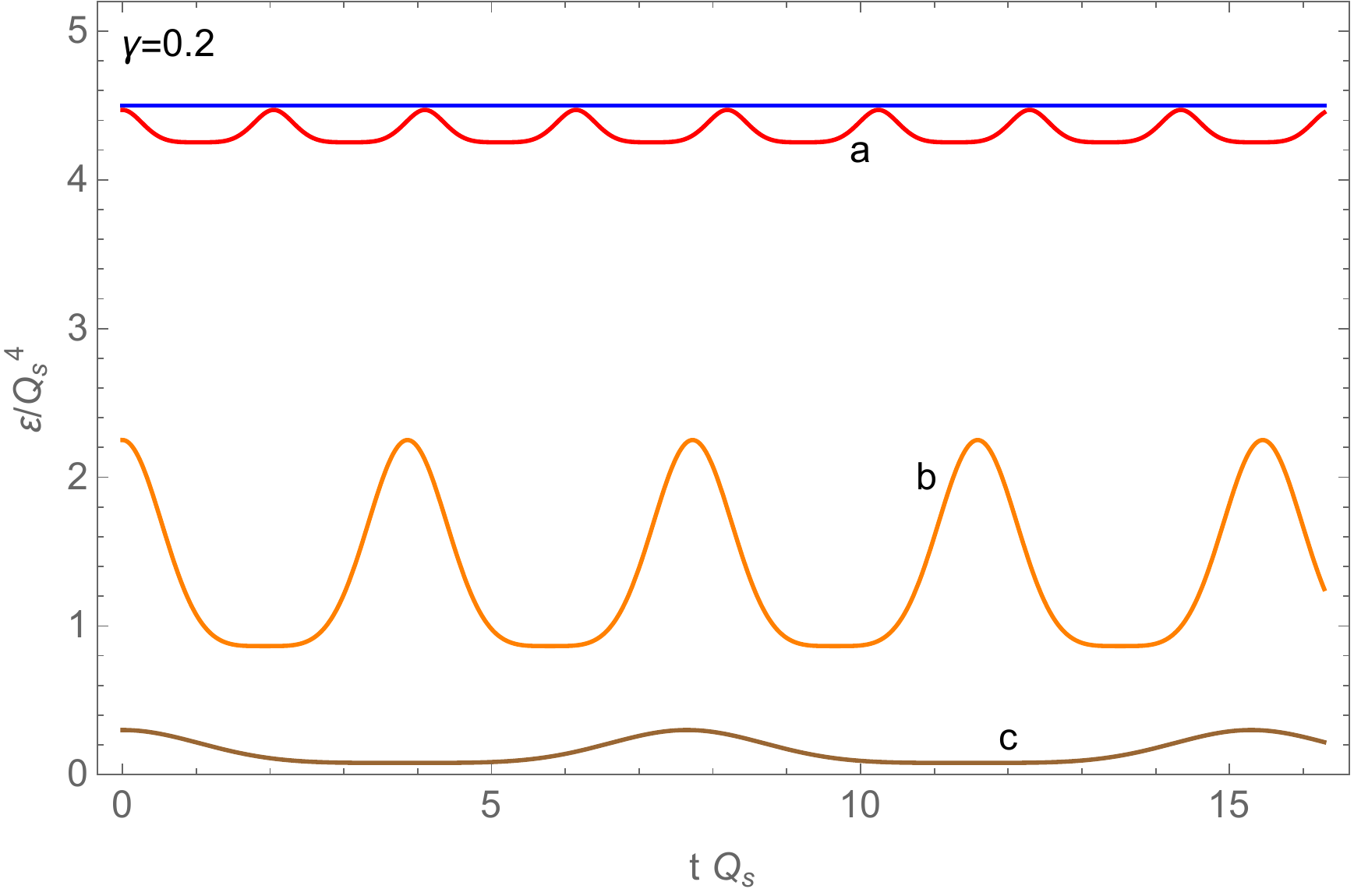}
\includegraphics[width=7cm,clip]{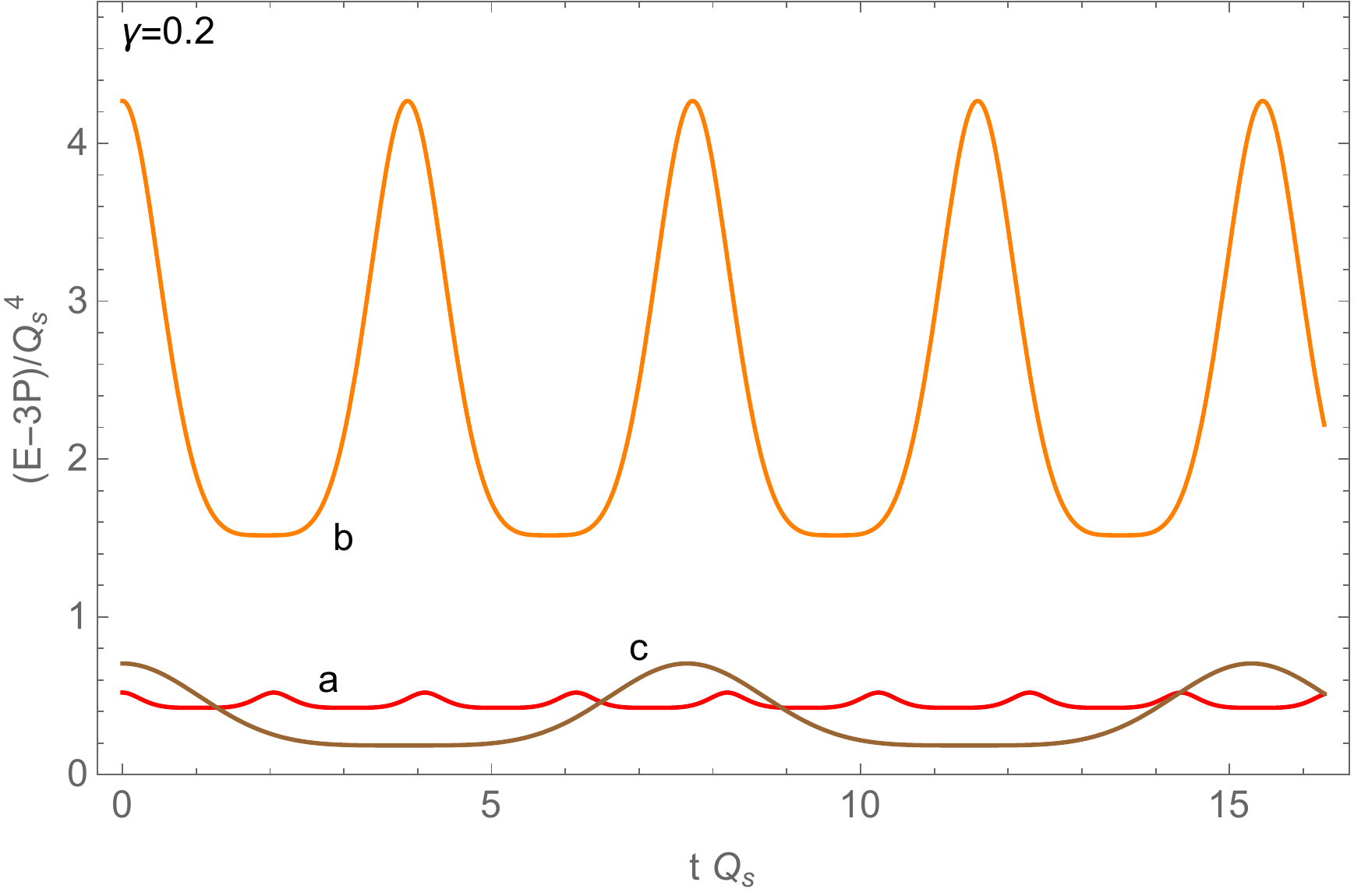}
\caption{Left panel: the Yang--Mills energy $\varepsilon/Q_s^4$ for the three cases (a) $p_0=1.49\ Q_s^4$, (b) $p_0=0.75\ Q_s^4$ and (c) $p_0=0.1\ Q_s^4$ with fixed total energy $\gamma E/Q_s^4=0.9$. Right panel: the interaction measure of the IR and the UV sector $E-3P$ for the same three sets of initial values.}
\label{fig:numericalsolution}       
\end{figure}

More concretely, the initial energy density $\varepsilon(0)$ for the YM sector and $\wE(0)$ for the soft sector take the values
\begin{eqnarray}
\varepsilon(0)&=&3p_0 \quad \mathrm{and}\\
\wE(0)&=&\frac{3N_c^2c}{8\pi^2}\frac{1}{\sqrt{1-3\frac{\gamma}{Q_s^4}p_0}\sqrt{1+\frac{\gamma}{Q_s^4}p_0}},
\end{eqnarray}
respectively in terms of the initial pressure $p_0$ in the YM sector and the mass parameter $c$ (of the pre-existing black hole). The total energy is found to be
\begin{equation}
E=\varepsilon(t)+\wE(t) \left(1- \frac{\gamma}{Q_s^4}  \varepsilon(t)\right)+\frac{3}{2}\frac{\gamma}{Q_s^4}\left(\wE(t) +\wP(t) \right)f'(t)^2=3p_{\ms{0}}+\frac{3N_c^2c}{8\pi^2}\sqrt{\frac{1-3\frac{\gamma}{Q_s^4}p_{\ms{0}}}{1+\frac{\gamma}{Q_s^4}p_{\ms{0}}}}.\label{eq:totalenergy}
\end{equation}
Note, that in order to obtain a regular $g^{\ts{(b)}}_{\mn}$ and thereby rendering the IR energy momentum tensor real, we have to impose a restriction on the parameter $\gamma$
\be
-\frac{Q_s^4}{p(t)}<\gamma<\frac{Q_s^4}{3p(t)}.\label{eq:condition1}
\ee
Furthermore in order to obtain regular solutions the coefficient of the second term in \eqref{eq:EOMiso} must not change sign, which imposes an additional restriction on $c$: for $\gamma$ satisfying (\ref{eq:condition1}) the said coefficient is positive at $t=0$ only if
\be
\frac{N_c^2 c}{2\pi^2Q_s^4}<\frac{2}{\vert\gamma\vert}\sqrt{1-3\frac{\gamma}{Q_s^4}p_0}\left(1+\frac{\gamma}{Q_s^4}p_0\right)^{3/2}.
\ee
However in the case $\gamma < 0$, the Yang--Mills pressure is initially in a local minimum. When the pressure reaches a local maximum, the total energy set by the choice of $p_0$, $c$ and $\gamma$ cannot not be matched if $N_c^2\gamma c/(2\pi^2Q_s^4)$ becomes too negative. One evaluates the left hand side of \eqref{eq:totalenergy} at 
$f(t_{\rm{max}})=f''(t_{\rm{max}})=0$, $f'(t_{\rm{max}})=\sqrt{2p_{\rm{max}}}$ and solves for $N_c^2\gamma c/(2\pi^2Q_s^4)$ the extremum of which has to be found numerically for each value of $\gamma p_0/Q_s^4$ in the range $(-1,0)$. This completes the discussion of the allowed region in $\{p_0, c\}$ parameter space of initial conditions in this toy example, which is marked by the shaded regions in Fig. \ref{fig:gammabounds}. There we also depict contours of constant values of $\gamma E/Q_s^4$. Note that lines lie completely within the allowed region provided that $-1<\gamma E/Q_s^4<1$.
For the discussion of the full numerical solution of Eq. \eqref{eq:EOMiso} with $\gamma=0.2$ we consider three different sets of $(p_0/Q_s^4,N_c^2c/2\pi^2Q_s^4)$ at fixed total energy $\gamma E/Q_s^4=0.9$. They are marked by three different shapes in Fig \ref{fig:gammabounds} with (a) $p_0=1.49\ Q_s^4$, (b) $p_0=0.75\ Q_s^4$ and (c) $p_0=0.1\ Q_s^4$. Case (a) corresponds to a large UV to IR energy ratio, in case (b) the ratio is of order unity and in case $(c)$ most of the energy resides in the IR. In Fig. \ref{fig:numericalsolution} we plot the time evolution of the Yang--Mills energy $\varepsilon/Q_s^4$ for the three cases in the first panel. The Yang--Mills energy exhibits oscillatory behaviour with increasing \textit{relative} amplitude as well as increasing wavelength as the UV-IR energy ratio decreases. The \textit{absolute} value of the amplitude is the largest when initially the energy is equally distributed among the IR and UV sectors. In this simple setup we cannot see dissipation of energy from the UV sector to the IR sector. However, as already mentioned we expect this to be the case when a finite value for $\beta$ is allowed and the three cases studied here might approximate three subsequent stages in the evolution of the system there.
In the second panel of Fig. \ref{fig:numericalsolution} we show the interaction measure of the IR and UV sectors defined by $E-3P$ for the same three cases. Note that this is positive only provided that $\gamma>0$.

The convergence of the iterative algorithm was shown to be very fast and reasonably stable. After four iterations there was no visible change left and after ten iterations both Eqs. \eqref{eq:EOMiso} and \eqref{eq:totalenergy} were satisfied to order $10^{-11}$. Only after several tens of iterations the accumulation of numerical errors became significant.

\section{Subleading quantum (kinetic) corrections}
At the subleading order, we need to add perturbative quantum corrections to the glasma. In the large $N_c$ limit, we do not need to account for quantum gravity corrections to the holographic nonperturbative sector. Assuming that we are dealing with very strong coupled dynamics for the IR, we can also ignore stringy higher derivative corrections to Einstein's gravity which will be otherwise necessary to account for departures from infinite 't Hooft coupling.

This semiholographic action with quantum corrections can be readily derived from first principles starting from the classical action (\ref{modified-glasma}) or (\ref{modified-glasma-2}). It is to be noted that the word classical refers to the glasma itself -- the holographic operators although coming from a strongly coupled quantum sector appear here only as self-consistent mean fields.

To this aim, we write the classical action (\ref{modified-glasma-2}) again as $S^{(0)}$ and just for sake of convenience we put $\alpha = \beta = 0$ (of course we can readily have non-zero values for these), i.e.
\begin{eqnarray}\label{glasma-ac-0}
S^{(0)}[A_\mu^a] &=& S_{\rm YM}[A_\mu^a] +\frac{1}{2} \int {\rm d}^4x \, \overline{\mathcal{T}}^{\mu\nu}g^{\rm (b)}_{\mu\nu} \nonumber\\
&=&S_{\rm YM}[A_\mu^a] + S^{\rm on-shell}_{grav}\Big[g^{\rm (b)}_{\mu\nu}= \eta_{\mu\nu} + \frac{\gamma}{Q_s^4} t^{\rm cl}_{\mu\nu}[A_\alpha^a]\Big],
\end{eqnarray}
with 
\begin{eqnarray}
t^{\rm cl}_{\mu\nu}[A_\alpha^a] = \frac{1}{N_c} {\rm tr} \left(F_{\mu\alpha}F_\nu^{\phantom{\mu}\alpha} - \frac{1}{4}\eta_{\mu\nu}F_{\alpha\beta}F^{\alpha\beta}\right).\label{source-12}
\end{eqnarray}
The quantum part (which can be derived using standard functional methods) is:
\begin{eqnarray}\label{glasma-ac-1}
S^{\rm (1)}[D_{\mu\nu}^{ab}, A_\alpha^c] &=& \frac{i}{2}{\rm Tr} \, \ln \, D^{-1} + \frac{i}{2}{\rm Tr}\left(D^{(0)-1}[A_\mu^a]D\right)\nonumber\\&&
+ S^{\rm on-shell}_{grav}\Big[g^{\rm (b)}_{\mu\nu} = \eta_{\mu\nu} + \frac{\gamma}{Q_s^4} \left(t^{\rm cl}_{\mu\nu}[A_\alpha^c] + t^{\rm q}_{\mu\nu}[D_{\rho\sigma}^{ab}]\right)\Big]\nonumber\\&&
-S^{\rm on-shell}_{grav}\Big[g^{\rm (b)}_{\mu\nu} = \eta_{\mu\nu} + \frac{\gamma}{Q_s^4} t^{\rm cl}_{\mu\nu}[A_\alpha^c] \Big],\nonumber\\
&=& \frac{i}{2}{\rm Tr} \, \ln \, D^{-1} + \frac{i}{2}{\rm Tr}\left(D^{(0)-1}[A_\mu^a]D\right)
\nonumber\\&&
+\frac{\gamma}{2Q_s^4} \int {\rm d}^4x \, \overline{\mathcal{T}}^{\mu\nu}t^{\rm q}_{\mu\nu}[D_{\rho\sigma}^{ab}],
\end{eqnarray}
where $\rm Tr$ denotes trace over colour, tensorial and Schwinger-Keldysh indices, and also integrations over the spacetime points, $D^{(0)-1}[A_\mu^a]$ is the inverse of the gluonic propagator in the classical $S_{\rm YM}[A_\mu^a]$ in presence of a background classical field configuration $A_\mu^a$ and
\begin{eqnarray}
t^{\rm q}_{\mu\nu}  & = &\frac{1}{4}\lim_{x\rightarrow y}\Big\{\Big[\partial_{\gamma}^{x}\partial^{y\gamma}\Big(\delta_{\mu}^{\alpha}\delta_{\nu}^{\beta}+\delta_{\nu}^{\alpha}\delta_{\mu}^{\beta}\Big)+\Big(\partial_{\mu}^{x}\partial_{\nu}^{y}+\partial_{\nu}^{x}\partial_{\mu}^{y}\Big)\eta^{\alpha\beta}\nonumber \\
 && -\partial_{\mu}^{x}\partial^{y\alpha}\delta_{\nu}^{\beta}-\partial^{x\alpha}\partial_{\mu}^{y}\delta_{\nu}^{\beta}-\partial^{x\alpha}\partial_{\nu}^{y}\delta_{\mu}^{\beta}-\partial_{\nu}^{x}\partial^{y\beta}\delta_{\mu}^{\alpha}\nonumber \\
 && -\eta_{\mu\nu}\Big[\eta^{\alpha\beta}\partial_{\gamma}^{x}\partial^{y\gamma}-\partial^{x\beta}\partial^{y\alpha}\Big]\Big\}{\rm tr}\Big(D_{\alpha\beta}(x,y)+D_{\alpha\beta}(y,x)\Big)\label{eq:tadpole-1}
\end{eqnarray}
is the one-loop tadpole contribution to the energy-momentum tensor arising from the quantum fluctuations.

\textit{The classical glasma fields $A_\mu^a$, the quantum fluctuations $D_{\mu\nu}^{ab}$ and the classical gravity solution giving nonperturbative dynamics should be solved together self-consistently in the perturbative 't Hooft coupling. The sources on the gravitational side include quantum fluctuations of the glasma fields.} 

\section{Towards a new description for collective flow}
It is useful to consider a simple limit for solving the quantum fluctuations. Let us assume that we are considering the final stage of QGP evolution where the classical fields $A_\mu^a$ have dissipated away so that we can ignore them. So, we can also set $t^{\rm cl}_{\mu\nu} = 0$, as a result of which \textit{at the leading order the boundary metric of the gravity solution is flat Minkowski space}. Nevertheless, \textit{at the leading order the soft holographic $\mathcal{T}^{\mu\nu}$ is not zero, but given by that of an expanding black hole.} For the moment, we ignore the expansion so that we consider that at late time we obtain a static thermal state. At the leading order, we can write
\begin{equation}
\overline{\mathcal{T}}^{(0)}_{\mu\nu}=\mathcal{T}^{(0)}_{\mu\nu} = \mathcal{T}^{\rm fluid}_{\mu\nu},
\end{equation}
where $ \mathcal{T}^{\rm fluid}_{\mu\nu}$ is the hydrodynamic energy-momentum tensor of a holographic fluid with appropriate transport coefficients. 

We can then readily see from (\ref{glasma-ac-1}) that 
\begin{equation}
D^{-1} = D^{(0)-1} - i\Sigma, \,\,\, {\rm with} \,\,\, \Sigma = \frac{\delta S^{\rm on-shell}_{grav}}{\delta D}.
\end{equation}
Explicitly in the Schwinger-Keldysh space,
\begin{eqnarray}
D^{(0)}_{\mu\nu}(p,p') &=&\mathcal{P}_{\mu\nu}\left(\begin{array}{cc}
\frac{1}{p^{2}+i\epsilon {\rm sgn}(p_{0})}-2\pi i\delta(p^{2})\theta(-p_{0}) & -2\pi i\theta(-p_{0})\delta(p^{2})\\
-2\pi i\theta(p_{0})\delta(p^{2}) & -\frac{1}{p^{2}-i\epsilon {\rm sgn}(p_{0})}-2\pi i\delta(p^{2})\theta(-p_{0})
\end{array}\right)\nonumber\\&&\delta^4(p-p'),
\end{eqnarray}
with
\begin{equation}
\mathcal{P}_{\mu\nu} = \eta_{\mu\nu}-\frac{p_{\mu}p_{\nu}}{p^{2}}.
\end{equation}
Also,
\begin{eqnarray}
\Sigma_{\mu\nu}(p,p^{\prime})  &=&-\frac{\gamma}{2Q_{s}^{4}}\Big[\eta_{\mu\nu}\mathcal{T}^{(0)\alpha\beta}(-p-p^{\prime})\,\,p_{\alpha}p_{\beta}^{\prime}-\mathcal{T}^{(0)}_{\mu\beta}(-p-p^{\prime})\,\,p_{\nu}p^{\prime\beta}\nonumber \\
 && \qquad\quad-\mathcal{T}^{(0)}_{\alpha\nu}(-p-p^{\prime})\,\,p^{\alpha}p^\prime_\mu+ \mathcal{T}^{(0)}_{\mu\nu}(-p-p^{\prime})\,\,p\cdot p^{\prime}\Big]\nonumber\\&&
 \times \left(\begin{array}{cc}
1 & 0\\
0 & -1
\end{array}\right).\label{eq:delsigma}
\end{eqnarray}
Note the change in $D$ by the self-energy term $\Sigma$ leads to a change in the boundary metric,
\begin{equation}
g^{\rm (b)}_{\mu\nu} = \eta_{\mu\nu} + g^{(1)}_{\mu\nu}, \,\,\, {\rm with}\,\,\,g^{(1)}_{\mu\nu} = \frac{\gamma}{Q_s^4} t_{\mu\nu}^{\rm q}[D],
\end{equation}
where $t_{\mu\nu}^{\rm q}[D]$ is given by (\ref{eq:tadpole-1}). Therefore, in order to preserve conservation of energy and momentum, i.e $\nabla_{\rm (b)\mu} \mathcal{T}^{\mu\nu} = 0$ with $\nabla_{(b)}$ being the covariant derivative constructed from $g^{(b)}_{\mu\nu}$, a small correction $\mathcal{T}^{(1)}_{\rm \mu\nu}$ should arise which should obey the conservation equations:
\begin{equation}
\partial_\mu \mathcal{T}^{(1)\mu\nu} = -\Gamma^{(1)\mu}_{\phantom{\mu}\mu \alpha} \mathcal{T}^{(0)\alpha\nu} -\Gamma^{(1)\nu}_{\phantom{\nu}\mu \alpha} \mathcal{T}^{(0)\mu\alpha},
\end{equation}
where
\begin{equation}
\Gamma^{(1)\mu}_{\phantom{\mu}\nu \rho} = \frac{1}{2}\eta^{\mu\alpha}\left(\partial_\nu g^{(1)}_{\alpha\rho} +\partial_\rho g^{(1)}_{\alpha\nu} - \partial_\alpha g^{(1)}_{\nu\rho}\right).
\end{equation}
Because the fluid-gravity correspondence follows from holography for any weakly curved boundary metric, we can consistently assume that $\mathcal{T}^{(1)}_{\mu\nu}$ also has a fluid form which however has forcing terms when viewed from the point of view of flat Minkowski background. We can show that we can absorb $\mathcal{T}^{(1)}_{\mu\nu}$ into $\mathcal{T}^{(0)}_{\mu\nu}$ which is a fluid in flat space by modifying the speed of sound and transport coefficients as functions of the velocity. So we get a specific generalisation of non-Newtonian fluid. We christen this as sesqui-hydrodynamics. It is to be noted that solving both sectors at the subleading order involves no iteration as we can take advantage of a systematic expansion. We will present further details and explicit results in the near future.
\section{Outlook}
We end with a list of questions that we should answer in the future.
\begin{enumerate}
\item How do the classical YM fields of the glasma thermalise with the dynamically formed black hole? Since the full system is stochastic (as a result of stochastic initial conditions), is the thermalisation process Markovian (as in Fokker-Planck system) or non-Markovian (with strong dependence on initial conditions)?
\item What kind of observables characterise the non-Markovian nature of thermalisation and how can we exploit them to learn more about nonperturbative dynamics of QCD?
\item Is the thermalisation process top-down, or bottom-up or riddled with quantum complexity? In the latter case what should be the experimental signatures?
\item How do we characterise the generalised non-Newtonian hydrodynamic collective flow of the combined hard-soft system at late time?
\end{enumerate}
At present we are only at the beginning of our explorations. Currently we are investigating all the above mentioned aspects.
 \begin{acknowledgement}
 We thank Yoshimasa Hidaka, Edmond Iancu, Aleksi Kurkela, Anton Rebhan, Stefan Stricker, Alexander Soloviev and Di-Lun Yang for collaborations on published and ongoing research works described in this proceedings. We thank Alexander Soloviev for comments on the manuscript. AM thanks the conveners of the Deconfinement session of the Confinement XII conference for invitation to give a talk. The research of AM is supported by a Lise Meitner fellowship of the Austrian Science Fund (FWF), project no. M1893-N27. The research of F. Preis is partially supported by the FWF project P26328-N27.
 \end{acknowledgement}

\begin{thebibliography}{8}
\bibitem{Iancu:2000hn} 
  E.~Iancu, A.~Leonidov and L.~D.~McLerran,
  Nucl.\ Phys.\ A {\bf 692}, 583 (2001)
  [hep-ph/0011241].
\bibitem{Gelis:2010nm} 
  F.~Gelis, E.~Iancu, J.~Jalilian-Marian and R.~Venugopalan,
  Ann.\ Rev.\ Nucl.\ Part.\ Sci.\  {\bf 60}, 463 (2010)
  [arXiv:1002.0333 [hep-ph]].
\bibitem{Kovner:1995ts}
A.~Kovner, L.D. McLerran, H.~Weigert, Phys. Rev. \textbf{D52}, 3809 (1995),
  \texttt{hep-ph/9505320}

\bibitem{Kovner:1995ja}
A.~Kovner, L.D. McLerran, H.~Weigert, Phys. Rev. \textbf{D52}, 6231 (1995),
  \texttt{hep-ph/9502289}

\bibitem{Lappi:2006fp}
T.~Lappi, L.~McLerran, Nucl. Phys. \textbf{A772}, 200 (2006),
  \texttt{hep-ph/0602189}
\bibitem{Kurkela:2011ti} 
  A.~Kurkela and G.~D.~Moore,
  JHEP {\bf 1112}, 044 (2011)
  [arXiv:1107.5050 [hep-ph]].
\bibitem{Kurkela:2014tea}
A.~Kurkela, E.~Lu, Phys. Rev. Lett. \textbf{113}, 182301 (2014),
  \texttt{1405.6318}

\bibitem{Kurkela:2015qoa}
A.~Kurkela, Y.~Zhu, Phys. Rev. Lett. \textbf{115}, 182301 (2015),
  \texttt{1506.06647}
 \bibitem{Policastro:2001yc} 
  G.~Policastro, D.~T.~Son and A.~O.~Starinets,
  Phys.\ Rev.\ Lett.\  {\bf 87}, 081601 (2001)
  [hep-th/0104066].

\bibitem{Back:2004je} 
  B.~B.~Back {\it et al.},
  Nucl.\ Phys.\ A {\bf 757}, 28 (2005)
  [nucl-ex/0410022].
   
  \bibitem{Iancu:2014ava}
E.~Iancu, A.~Mukhopadhyay, JHEP \textbf{06}, 003 (2015), \texttt{1410.6448}


\bibitem{Mukhopadhyay:2015smb} 
  A.~Mukhopadhyay, F.~Preis, A.~Rebhan and S.~A.~Stricker,
  JHEP {\bf 1605}, 141 (2016)
  [arXiv:1512.06445 [hep-th]].
  
  \bibitem{deHaro:2000vlm} 
  S.~de Haro, S.~N.~Solodukhin and K.~Skenderis,
  Commun.\ Math.\ Phys.\  {\bf 217}, 595 (2001)
  doi:10.1007/s002200100381
  [hep-th/0002230].
  
  \bibitem{Chesler:2008hg} 
  P.~M.~Chesler and L.~G.~Yaffe,
  Phys.\ Rev.\ Lett.\  {\bf 102}, 211601 (2009)
  [arXiv:0812.2053 [hep-th]].
  
\bibitem{PhysRevD.16.1712}
L.S. Brown, J.P. Cassidy, Phys. Rev. D \textbf{16}, 1712 (1977)

\bibitem{Herzog:2013ed}
C.P. Herzog, K.W. Huang, Phys. Rev. \textbf{D87}, 081901 (2013),
  \texttt{1301.5002}



\end{thebibliography}

%
%


%
%

\end{document}